\documentclass[sn-mathphys,Numbered]{sn-jnl}

\usepackage{graphicx}%
\usepackage{multirow}%
\usepackage{amsmath,amssymb,amsfonts}%
\usepackage{amsthm}%
\usepackage{mathrsfs}%
\usepackage[title]{appendix}%
\usepackage{xcolor}%
\usepackage{textcomp}%
\usepackage{manyfoot}%
\usepackage{booktabs}%
\usepackage{algorithm}%
\usepackage{algorithmicx}%
\usepackage{algpseudocode}%
\usepackage{caption}
\usepackage{listings}%
\usepackage{wrapfig}%

\usepackage{scalerel}
\usepackage{tikz}
\usepackage{tikz-cd}

\usetikzlibrary{positioning}
\usetikzlibrary{svg.path}

\definecolor{orcidlogocol}{HTML}{A6CE39}
\tikzset{
  orcidlogo/.pic={
    \fill[orcidlogocol] svg{M256,128c0,70.7-57.3,128-128,128C57.3,256,0,198.7,0,128C0,57.3,57.3,0,128,0C198.7,0,256,57.3,256,128z};
    \fill[white] svg{M86.3,186.2H70.9V79.1h15.4v48.4V186.2z}
                 svg{M108.9,79.1h41.6c39.6,0,57,28.3,57,53.6c0,27.5-21.5,53.6-56.8,53.6h-41.8V79.1z M124.3,172.4h24.5c34.9,0,42.9-26.5,42.9-39.7c0-21.5-13.7-39.7-43.7-39.7h-23.7V172.4z}
                 svg{M88.7,56.8c0,5.5-4.5,10.1-10.1,10.1c-5.6,0-10.1-4.6-10.1-10.1c0-5.6,4.5-10.1,10.1-10.1C84.2,46.7,88.7,51.3,88.7,56.8z};
  }
}

\newcommand\orcidicon[1]{\href{https://orcid.org/#1}{\mbox{\scalerel*{
\begin{tikzpicture}[yscale=-1,transform shape]
\pic{orcidlogo};
\end{tikzpicture}
}{|}}}}

\theoremstyle{thmstyleone}%
\newtheorem{theorem}{Theorem}
\theoremstyle{thmstyletwo}%
\newtheorem{remark}{Remark}%

\theoremstyle{thmstylethree}%
\newtheorem{definition}{Definition}%

\begin{document}

\title[The Traveling Mailman: Topological Optimization Methods for User-Centric Redistricting]{The Traveling Mailman: Topological Optimization Methods for User-Centric Redistricting}


\author{\fnm{Nelson} \sur{Col\'on Vargas} \orcidicon{0009-0009-9038-7328}}


\affil{\orgdiv{School of International and Public Affairs}, \orgname{Columbia University}} 



\abstract{This study introduces a new districting approach using the US Postal Service network to measure community connectivity. We combine Topological Data Analysis with Markov Chain Monte Carlo methods to assess district boundaries' impact on community integrity. Using Iowa as a case study, we generate and refine districting plans using KMeans clustering and stochastic rebalancing. Our method produces plans with fewer cut edges and more compact shapes than the official Iowa plan under relaxed conditions. The low likelihood of finding plans as disruptive as the official one suggests potential inefficiencies in existing boundaries. Gaussian Mixture Model analysis reveals three distinct distributions in the districting landscape. This framework offers a more accurate reflection of community interactions for fairer political representation.}

\keywords{Redistricting, Gerrymandering, Voting Rights, User-Experience, Topological Data Analysis, Markov Chain Monte Carlo}



\maketitle

\pagebreak

\tableofcontents
\newpage 


\section{Introduction}\label{sec1}

Electoral districting is crucial for shaping political representation and resource distribution, but it often faces scrutiny over issues such as gerrymandering and inadequate representation. Traditional redistricting practices have primarily focused on population equality and political boundaries, often resulting in districts that fail to accurately reflect community cohesion or interests. DeFord, Duchin, and Solomon \cite{deford2019mcmc} introduced a new method for creating districts by cutting edges of the adjacency graph---with the idea that minimal cuts make for better maps---and then modeling plans by observing the distribution of the original plan's cut edges versus the samples. However, this approach doesn't fully capture the true spatial distribution of populations within counties and assigns equal weight to all cuts, regardless of their impact on community connectivity.

Our method builds upon and enhances the work of DeFord et al. by utilizing a graph that captures more than just proximity among districts. We use post offices as representatives of population hubs, providing a multidimensional representation of population distribution within counties or tracts. The postal network, optimized for equal access and connections, serves as a proxy for community cohesiveness. By employing persistent homology within Topological Data Analysis (TDA), we leverage detailed information about spatial distribution and connections within the population. This approach captures nuances beyond mere population counts; moving the same county can have varying disruptive effects depending on where the cut occurs, as edges are not uniformly distributed. Our analysis of how districting plans intersect with the postal network allows for a more sophisticated evaluation of community cohesion and integrity, surpassing traditional centroid-based measures. This method provides insights into the impact of redistricting on community structures, offering a more nuanced assessment than conventional approaches.

  \begin{figure}[h]
  \begin{minipage}{0.5\textwidth}
    \begin{center}
     \captionsetup{justification=centering}
    \includegraphics[width=\linewidth]{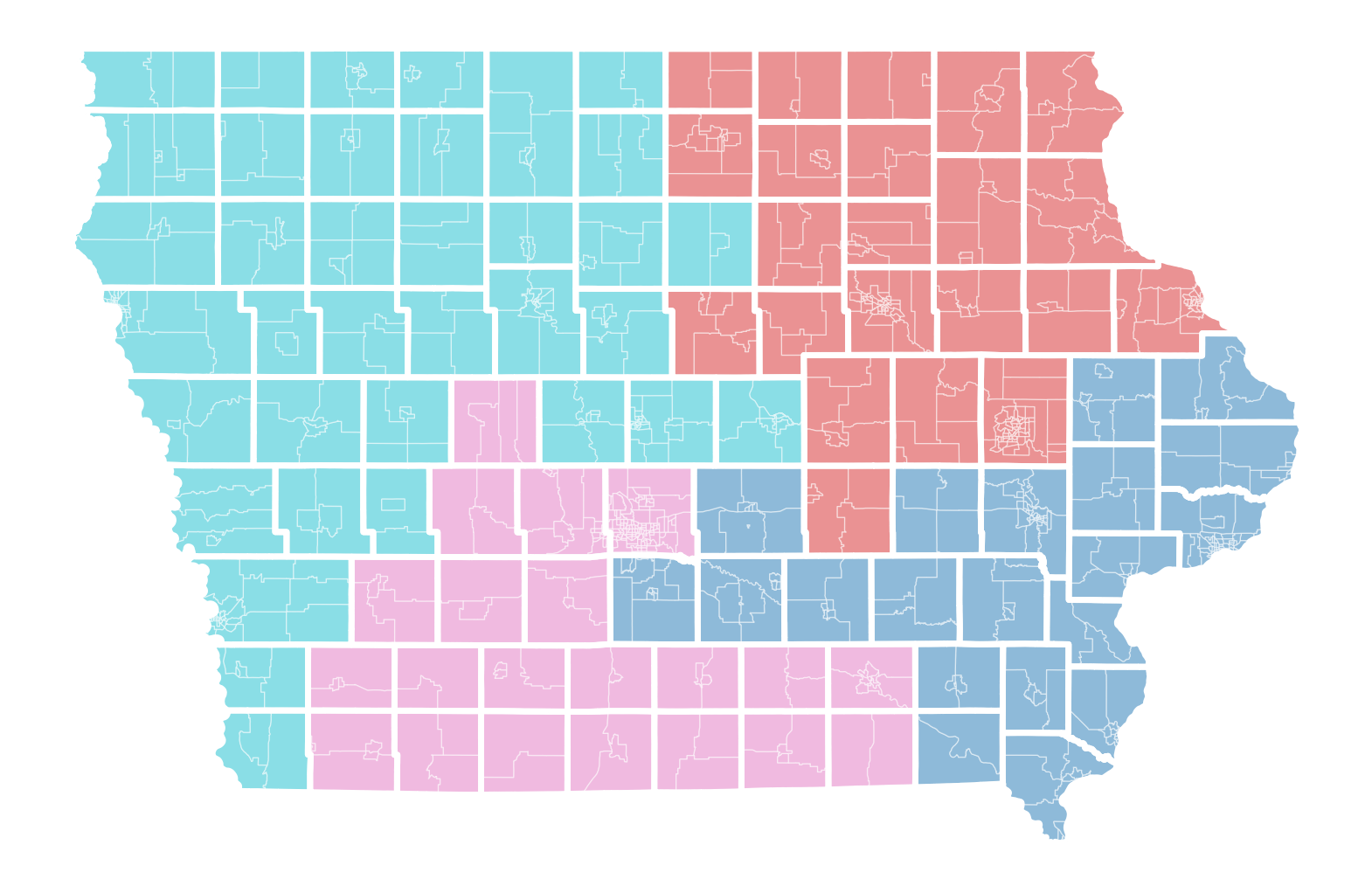}
    \caption{Iowa Districts (2021)}
    \end{center}
    \end{minipage}\hfill
      \begin{minipage}{0.5\textwidth}
    \begin{center}
     \captionsetup{justification=centering}
    \includegraphics[width=\linewidth]{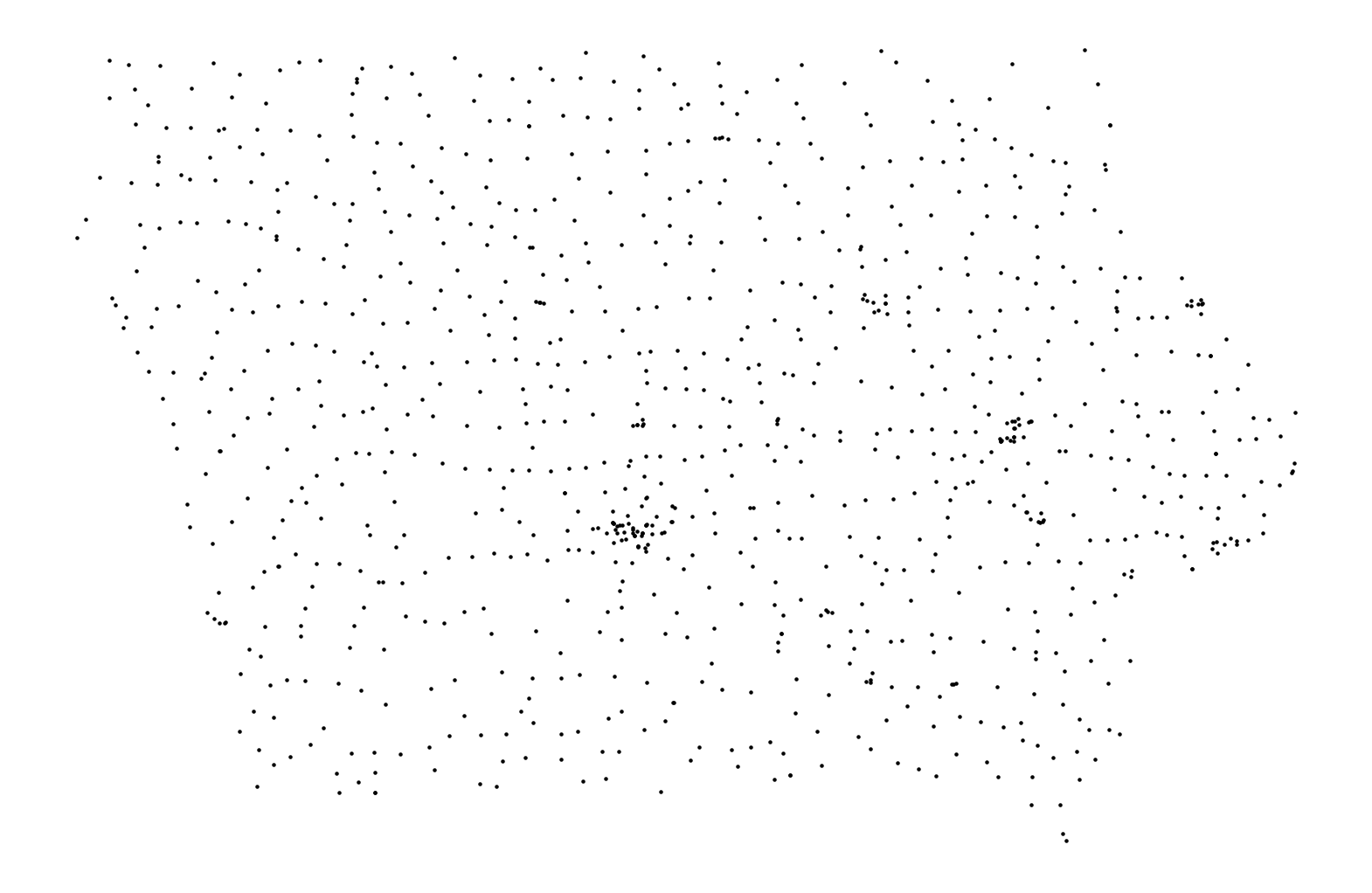}
    \caption{Iowa Post Offices}
    \label{networkonly100}
    \end{center}
    \end{minipage}\hfill
\end{figure}

We draw inspiration from an unconventional source: the Waffle House Index used by the Federal Emergency Management Agency (FEMA) \cite{wafflehouseindex}. An informational metric that uses the operational status of Waffle House restaurants as an indicator of how severely a community has been affected by a natural disaster. In a similar vein, we propose using the postal network as a proxy for community structure.

\begin{figure}[h]
    \begin{minipage}[t]{0.5\textwidth}
        \begin{center}
            \captionsetup{justification=centering}
            \includegraphics[width=.95\linewidth]{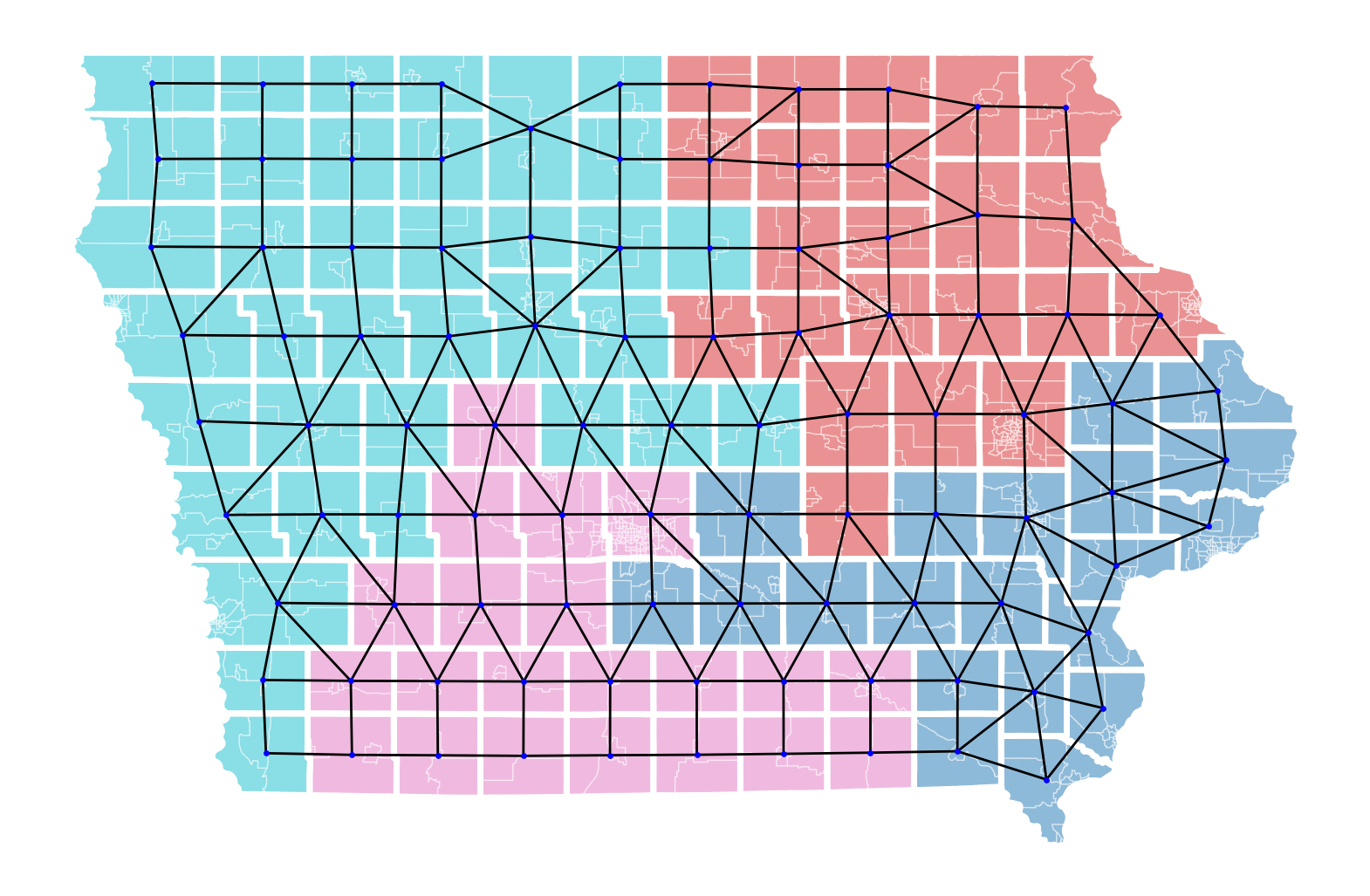}
            \caption{Iowa Districts with Adjacency Graph}
            \label{iowa_districts_adjacency}
        \end{center}
        \vspace{0pt}
    \end{minipage}\hfill
    \begin{minipage}[t]{0.5\textwidth}
        \begin{center}
            \captionsetup{justification=centering}
            \includegraphics[width=\linewidth]{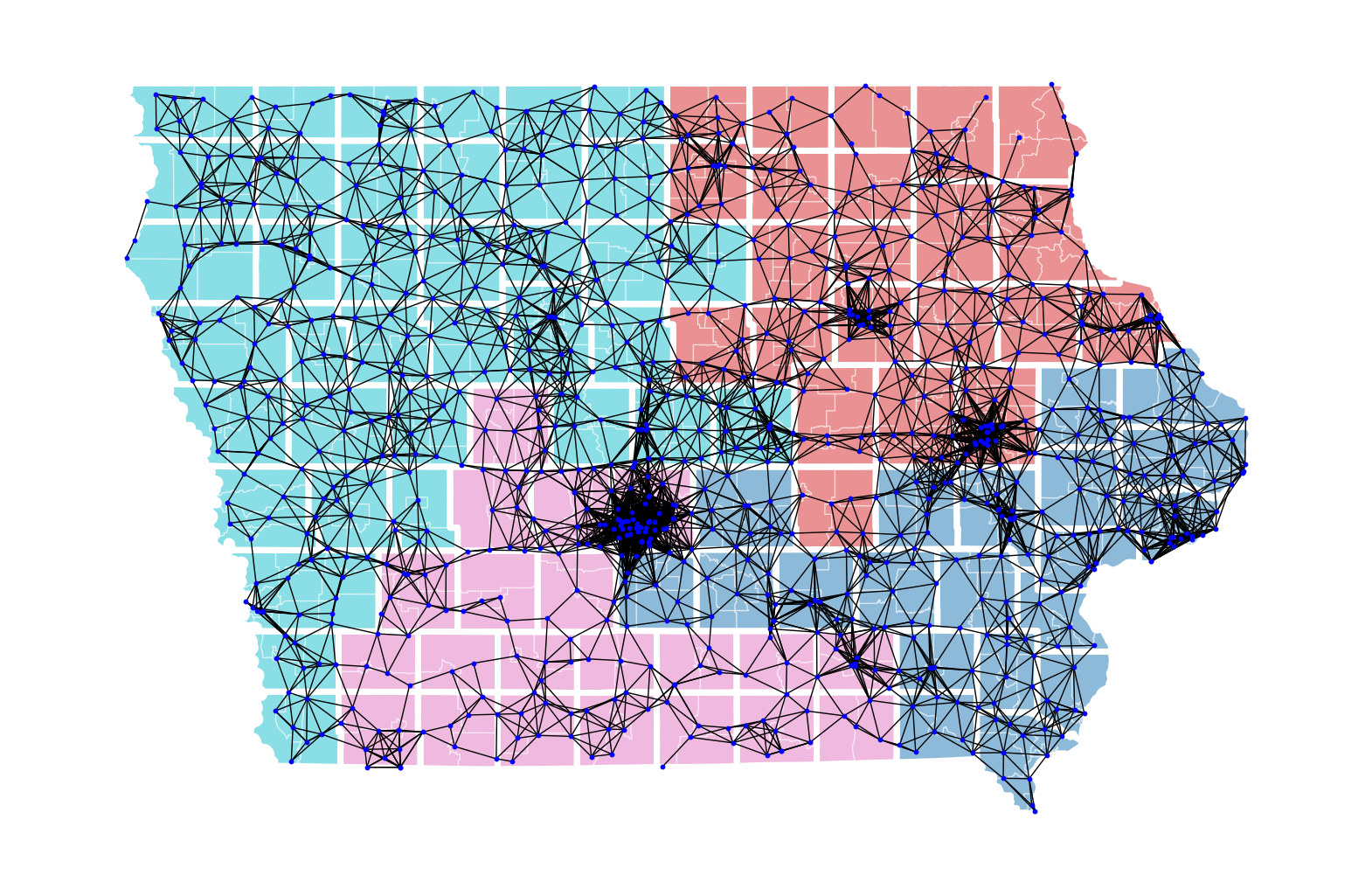}
            \caption{Iowa Districts with TDA Postal Network $(\epsilon = 14mi)$}
            \label{iowa_districts_network}
        \end{center}
        \vspace{0pt}
    \end{minipage}
\end{figure}

Similar to DeFord et al., we model districting plans through Markov Chain Monte Carlo (MCMC) methods. However, we evaluate the cut edges of the persistent homology graph rather than a simple adjacency graph, allowing for a more comprehensive analysis of the impact on community structures. Our method generates a large number of districting maps that meet selected criteria using MCMC techniques. By generating these maps randomly and ensuring they satisfy the predefined criteria, we can use statistical methods to assess the likelihood that an existing districting map could have arisen by chance. This probabilistic evaluation helps to determine if the existing map is an outlier or within the range of plausible districting configurations. By combining the concept of cut edges with our postal network representation and TDA, we can quantitatively assess the preservation of community structures in proposed districting plans.

Like many other researchers in redistricting (\cite{deford2019mcmc, mccartan2023pareto}), we used Iowa as a case study due to its requirement to make districts respect county lines, simplifying computations. However, this process could be extended to the Census tract level with more computational power.

\subsubsection*{Key Contributions}
\begin{itemize}
    \item \textbf{Introduction of SRKMeans Districts:} Developed the Stochastic Rebalancing KMeans (SRKMeans) for initializing district searches.
    \item \textbf{Integration of TDA with Redistricting:} Utilized Persistent Homology to quantitatively capture community structures across scales, enhancing the redistricting process.
    \item \textbf{Novel Use of Postal Network:} Employed the postal network to mirror community interactions, providing a practical and efficient proxy for community focal points.
    \item \textbf{Extension of MCMC-based District Plan Evaluation:} Built upon the work of DeFord et al. by evaluating how the generated district plans intersect with the postal network, using cut edges in the postal network as a measure of community integrity.
    \item \textbf{Optimization Method:} Developed a method to find better configurations within set parameters, effectively minimizing cut edges.
\end{itemize}

\subsubsection*{Data Collection and Code}
For the pseudocode of the algorithms used in this manuscript refer to the appendix. All the data, and code is available on this GitHub repository: \url{https://github.com/nelabdiel/TDARedistricting}. 

\section{Theoretical Framework and Definitions}

\subsection{Districting}

\begin{definition}
An \textbf{electoral district} is a geographic area represented by a legislator, delineated to organize the election of representatives. District boundaries typically consider factors like population size, geographic continuity, and the cohesion of community interests.
\end{definition} \vspace{10pt}

\noindent Let $S$ be an arbitrary state in the US. The number of districts assigned to $S$ is determined by the formula: 
\begin{equation}
N \approx 435 \times \frac{Population_{S}}{Population_{USA}}
\end{equation} \vspace{10pt}

\begin{definition}
\label{def:general_districting}
A collection of districts, $\Delta$, for any given state $S$, is a $N$-partition of $S$ that satisfies the following conditions:
\begin{itemize}
    \item Equal population: i.e., for any two districts $\delta_{i}, \delta_{j} \in \Delta$, it holds that $\|\delta_{i}\| \approx \| \delta_{j}\|$
    \item Contiguity: Each district must be a single connected component without exclaves.
    \item Compactness: The shape of the district should avoid unnecessary elongation or division, adhering as closely as possible to conventional geometric shapes.
\end{itemize}
\end{definition} \vspace{10pt}

\begin{definition}[Polsby-Popper Score \cite{polsby1991compactness}]
Let $\delta$ be a district, the \textbf{Polsby-Popper score test} defined as
\begin{equation}
PP(\delta) = \frac{4\pi \left|\delta\right|}{|\partial \delta|^2},
\end{equation}
where $\left|\delta\right|$ represents the area of $\delta$, and $|\partial \delta|$ denotes the perimeter (length) of the boundary of $\delta$. 
\end{definition} \vspace{10pt}

\begin{remark}
The Polsby-Popper score compares the area of the district to the area of a circle with the same boundary length, thus providing a measure of how close the district is to being circular, the shape with maximal compactness. A circle achieves the highest possible Polsby-Popper score of 1, indicating perfect compactness, while lower scores suggest more elongated or irregular boundaries.
\end{remark}  \vspace{10pt}

Recognizing the complexities inherent in redistricting, we focus on a relaxation of the districting problem and Definition~\ref{def:general_districting} where controlled deviations are permitted, based on individual preference and tolerance. This approach allows for more flexible responses to geographical and demographic challenges while maintaining the integrity of districting principles.

\begin{definition}
A \textbf{districting plan} is \textit{admissible} if it satisfies the following conditions:
\begin{enumerate}
\item \textbf{Population Balance:} The population of each district must be within a specific deviation threshold of the average district population. Mathematically, for any district $\delta_i$ in the plan $\Delta$:
\begin{equation}
\left| P(\delta_i) - P_{\text{avg}} \right| \leq \theta \cdot P_{\text{avg}},
\end{equation}
where $P_{\text{avg}}$ is the state total population divided by the number of districts and $\theta$ is the maximum allowed deviation.
\item \textbf{Contiguity:} A district $\delta_i$ is contiguous if it is topologically connected, meaning it consists of a single piece without any disjoint or isolated parts. 
\item \textbf{Compactness:} Each district's shape must meet a minimum compactness criterion defined by the Polsby-Popper measure. For a district $\delta_i$:
\begin{equation}
PP(\delta_i) \geq (1 - \kappa) \cdot PP_{\text{min}},
\end{equation}
where $PP(\delta_i)$ is the Polsby-Popper measure of district $\delta_i$, $PP_{\text{min}}$ is the measure of the least compact district in the existing plan, and $\kappa$ is the minimum acceptable compactness relative to $PP_{\text{min}}$.

\end{enumerate}
\end{definition}  \vspace{10pt}

\begin{remark}
The \textbf{population deviation threshold} $\theta$ is typically set based on legal and practical considerations to ensure fairness and equality in representation. For instance, a $\theta$ value of 0.05 indicates a tolerance of 5\% deviation from the average district population.
\end{remark} \vspace{10pt}

\begin{remark}
\textbf{Geospatial contiguity} in districting ensures that every part of a district is reachable without crossing the district boundary. In practical terms, this means that the district should not be comprised of multiple disjoint parts.
\end{remark} \vspace{10pt}

\begin{remark}
The \textbf{compactness relaxation factor} $\kappa$ is set to accommodate natural and civic boundaries which may necessitate less compact district shapes. $\kappa$ set to 0.05, a tolerance of no less than 95\% as compact as the least compact existing district.
\end{remark} \vspace{10pt}

  \begin{figure}[h]
    \begin{center}
     \captionsetup{justification=centering}
    \includegraphics[width=.9\linewidth]{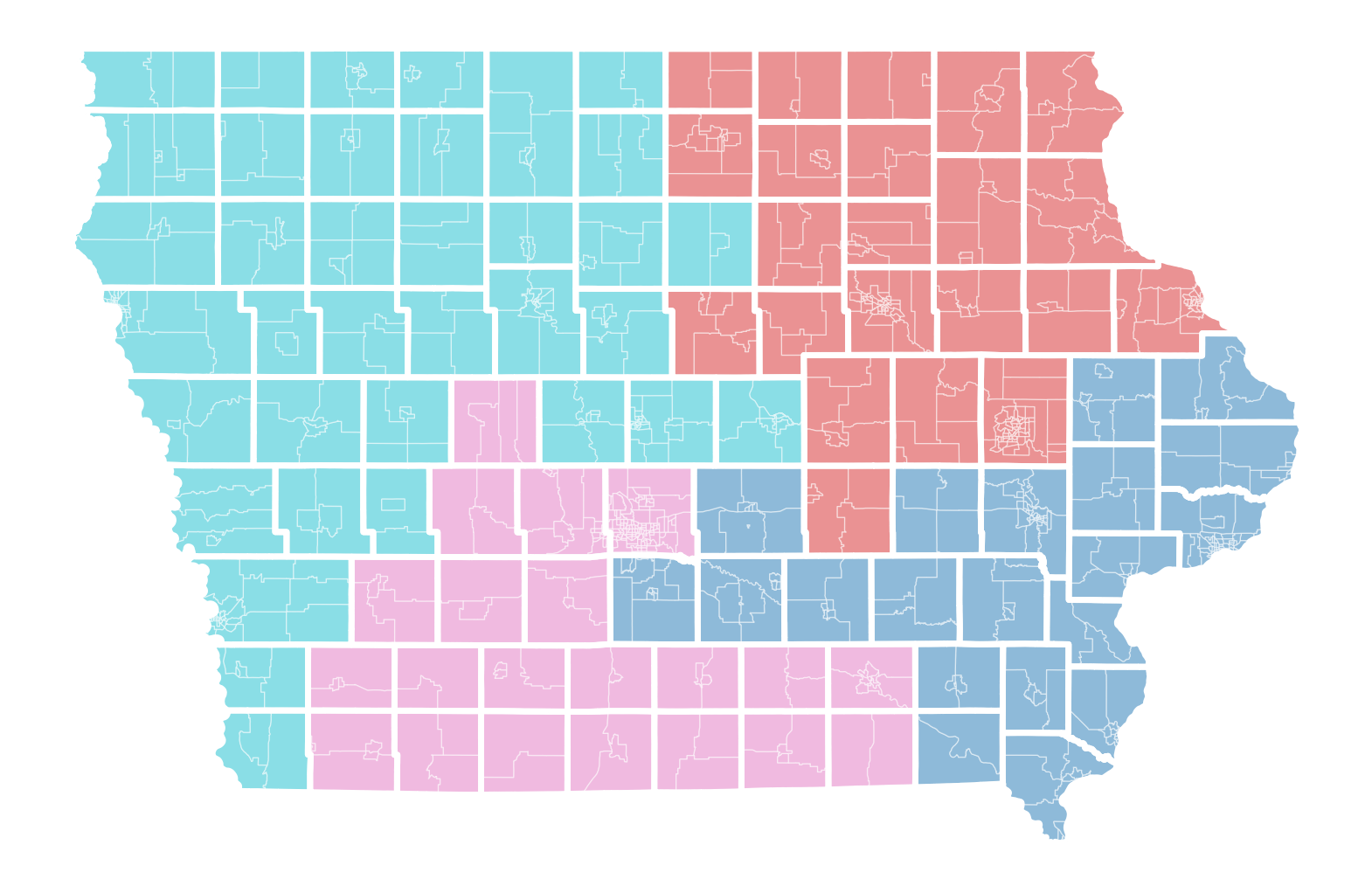}
    \caption{Iowa's Districting Map.}
    \end{center}
\end{figure}

\begin{table}[h]
    \centering
    \begin{tabular}{|c|c|c|c|}
        \hline
        \textbf{District} & \textbf{Population} & \textbf{Population Deviation (\%)} & \textbf{Polsby Popper} \\ \hline
        0 & 797,584 & -0.0010 & 0.27 \\ \hline
        1 & 797,589 & -0.0004 & 0.39 \\ \hline
        2 & 797,551 & -0.0052 & 0.32 \\ \hline
        3 & 797,645 &  { }0.0066 & 0.26 \\ \hline
    \end{tabular}
    \caption{Iowa Districts Info}
    \label{table:district_population}
\end{table}

\subsection{Topological Data Analysis}

\begin{definition}[Topological Data Analysis] 
Topological Data Analysis (TDA) is a framework for analyzing data using concepts from topology and geometry. Given a dataset $X$ embedded in a high-dimensional space, TDA seeks to describe the topological structure of $X$ through its simplicial complexes, built at multiple scales to capture connectivity, holes, and other geometric features that are invariant under continuous deformations.
\end{definition} \vspace{10pt}

\begin{definition}[Persistent Homology] 
Persistent Homology provides a multi-scale topology of a data set by constructing a series of nested subspaces $\{X_t\}_{t \in \mathbb{R}}$, filtered by a parameter $t$. It captures the birth and death of topological features as $t$ varies, formalizing the persistence of these features. If $H_k(X_t)$ denotes the $k$-th homology group of the space $X_t$, then the persistent homology groups are given by:
\begin{equation*}
PH_k(X) = \{ (b,d) \in \mathbb{R}^2 : b < d \text{ and } \text{homology class } [c] \text{ is born at } b \text{ and dies at } d \},
\end{equation*}
where $[c]$ represents a homology class in $H_k(X_t)$. The interval $(b, d)$ is called the persistence interval of the class $[c]$, capturing its longevity across the filtration.
\end{definition} \vspace{10pt}

\begin{figure}[h]
    \begin{center}
    \captionsetup{justification=centering}
        \includegraphics[width=.6\linewidth]{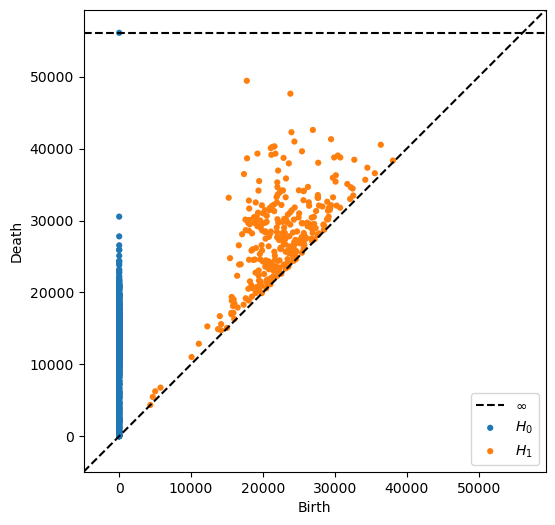}
        \caption{Homology of Iowa's Postal Network}
    \end{center}
\end{figure}

Persistence in topological data analysis measures the lifespan of features in a dataset across different scales. To establish a significant threshold for persistence, this study employs a percentile-based method. For a persistence diagram that records the birth and death times of topological features, the persistence value for a feature is calculated as the difference between its death and birth times.
Given a set of persistence values $\{p_i\}$ from the persistence diagram, the threshold $\epsilon$ is defined at a specific percentile $p$. This threshold helps in distinguishing significant topological features from noise. Mathematically, the threshold $\epsilon$ is determined as follows:
\[
\epsilon = \text{Quantile}(\{p_i\}, p)
\]
where $p$ represents the chosen percentile, typically adjusted to balance sensitivity and specificity in the analysis.

  \begin{figure}[h]
  \begin{minipage}{0.5\textwidth}
    \begin{center}
     \captionsetup{justification=centering}
    \includegraphics[width=\linewidth]{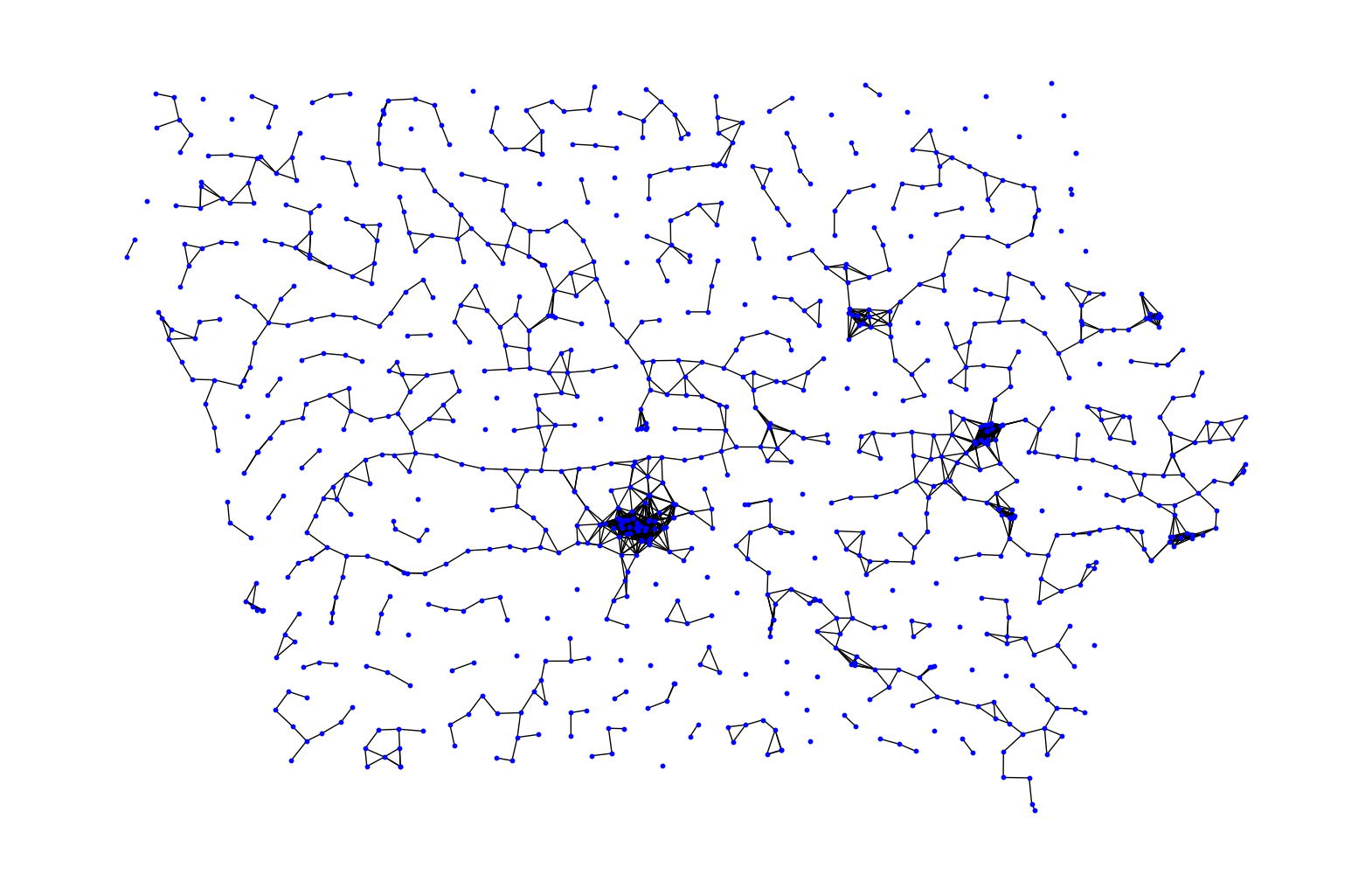}
    \caption{Iowa's Postal Network. $p=90\%$, $\epsilon \approx 8mi$.}
    \end{center}
    \end{minipage}\hfill
      \begin{minipage}{0.5\textwidth}
    \begin{center}
     \captionsetup{justification=centering}
    \includegraphics[width=\linewidth]{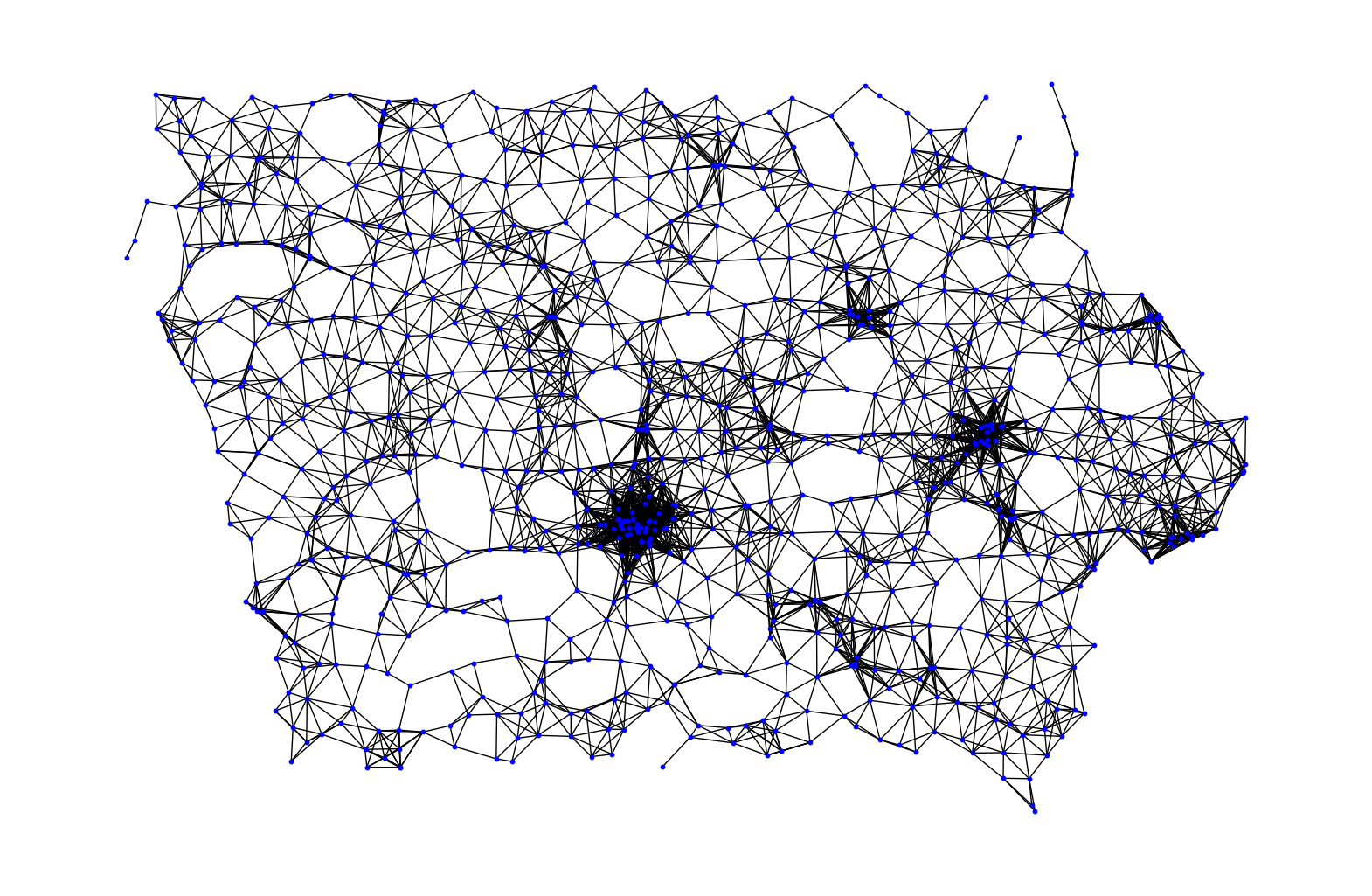}
    \caption{Iowa's Postal Network. $p=100\%$, $\epsilon \approx 14mi$.}
    \label{networkonly100}
    \end{center}
    \end{minipage}\hfill
\end{figure}

\begin{remark}[Choice of Percentile]
The choice of percentile $p$ significantly influences the topological analysis. A higher percentile (e.g., 95th) ensures that only the most persistent features are considered, minimizing the influence of transient noise. This method is particularly useful in datasets with variable or unknown noise levels, providing a robust means to ensure the reliability of derived topological insights.
\end{remark}\vspace{10pt}

\begin{definition}
Let $\mathcal{G}$ be a graph constructed from the network of postal offices within a geographic region, where vertices represent postal offices and edges represent the paths connecting them. A districting plan $\Delta$ is said to preserve community integrity if the higher-dimensional homological features of the graph remain minimally disrupted and the induced subgraphs of $\mathcal{G}$ corresponding to each district maintain a high level of connectivity exhibiting minimal edge cuts relative to the original graph $\mathcal{G}$, aiming to preserve the overall structure and connectivity of the community network.
\end{definition} \vspace{10pt}

\subsection{KMeans Clustering}

\begin{definition}[KMeans Clustering] 
\textbf{KMeans Clustering} is a method of vector quantization that aims to partition $n$ observations into $k$ clusters, in which each observation belongs to the cluster with the nearest mean, serving as a prototype of the cluster. Formally, given a set of data points ${x_1, x_2, \ldots, x_n}$, the objective is to minimize the within-cluster sum of squares:
\begin{equation}
\min_{{C_i}} \sum_{i=1}^{k} \sum_{x \in C_i} | x - \mu_i |^2,
\end{equation}
where $C_i$ is the set of points in cluster $i$ and $\mu_i$ is the mean of points in $C_i$.
\end{definition} \vspace{10pt}

\begin{remark}
KMeans Clustering is particularly effective for partitioning data into distinct groups, making it a useful tool for initializing district plans in redistricting problems. By clustering geographic centroids weighted by population, we ensure that the initial districting configuration respects population balance and geographic proximity.
\end{remark} \vspace{10pt}

\begin{theorem}[Convergence of KMeans \cite{bottou1994kmeans}]
Given a set of $n$ data points in $d$ dimensions, the KMeans algorithm iteratively improves cluster assignments and mean positions, converging to a local minimum of the within-cluster sum of squares. Although the global minimum is not guaranteed, the convergence to a local minimum ensures that the algorithm provides a reasonably optimal partitioning of the data.
\end{theorem} \vspace{10pt}

\begin{remark}
In this study, we apply KMeans Clustering to the centroids of counties, weighted by population, to create an initial districting plan. This initial plan is then refined using stochastic rebalancing techniques and evaluated based on the criteria of population balance, contiguity, and compactness.
\end{remark} \vspace{10pt}

This subsection introduces KMeans Clustering, provides a formal definition, and outlines its application in the context of redistricting, setting the stage for its use in generating initial district plans.

\subsection{Markov Chain Monte Carlo}

\begin{definition}[Markov Chain Monte Carlo] 
\textbf{Markov Chain Monte Carlo (MCMC)} is a class of algorithms that sample from a probability distribution based on constructing a Markov chain that has the desired distribution as its equilibrium distribution. The state of the chain after a large number of steps is then used as a sample of the desired distribution.
\end{definition} \vspace{10pt}

\begin{theorem}[Convergence of Markov Chain Monte Carlo \cite{bishop2006ml}]
Let $\{X_n\}_{n=0}^{\infty}$ be a Markov chain on a state space $\mathcal{X}$ with transition probabilities satisfying detailed balance relative to a probability distribution $\pi$ on $\mathcal{X}$. Then, $\pi$ is a stationary distribution of the Markov chain, and for any measurable function $f: \mathcal{X} \rightarrow \mathbb{R}$, under certain regularity conditions, the following convergence in distribution holds:
\begin{equation}
\lim_{n \to \infty} P(X_n \in A) = \pi(A) \text{ for all measurable sets } A \subset \mathcal{X},
\end{equation}
where $X_n$ denotes the state of the Markov chain at step $n$.
\end{theorem} \vspace{10pt}

\subsubsection{Rationale for Using MCMC}

Markov Chain Monte Carlo (MCMC) methods are employed to generate a wide variety of admissible districting plans. This approach allows us to estimate the posterior distribution of the number of cut edges across these plans. By sampling from this distribution, MCMC facilitates a robust statistical analysis of how likely any given plan is to occur under a set of fairness and balance criteria predefined by the model. This is pivotal in assessing the fairness of an initial plan by comparing it against this posterior distribution.

\begin{definition}
The \textbf{posterior distribution} in the context of this study refers to the distribution of the number of cut edges after observing the data from numerous MCMC iterations. This distribution reflects the likelihood of various districting outcomes given the constraints and conditions set by the model, such as population equality, contiguity, and compactness.
\end{definition}

By exploring the posterior distribution, we can statistically evaluate how the original districting plan compares with randomly generated plans that meet legal and fairness criteria. If the original plan has a significantly low probability of occurrence in this distribution, it may suggest potential issues with how the districts were delineated, such as gerrymandering or bias.

\section{Methodology}\label{methodology}
The proposed method employed an MCMC approach to iteratively generate and evaluate districting plans. This process involved several steps: initialization, proposal generation, acceptance criteria, and recording of results. The primary goal was to create districting plans that adhered to specified criteria while exploring the space of possible configurations to assess community integrity through minimal cut edges.

\subsection{Choice of $p$ for this Study}

For this study, we selected the 100th percentile ($p=100\%$), corresponding to an epsilon of approximately 14 miles. This threshold was chosen to ensure that we capture the entire range of topological features, providing a comprehensive view of the community structure through the postal network. This approach helps in thoroughly understanding the connectivity and potential disruptions caused by districting plans. However, the choice of the 100th percentile also means that we consider even the shortest-lived features, which might include noise. The decision to use this threshold was based on the goal of achieving a detailed analysis and ensuring that no significant community structures are overlooked.

  \begin{wrapfigure}{r}{0.45\textwidth}
    \begin{center}
     \captionsetup{justification=centering}
    \includegraphics[width=\linewidth]{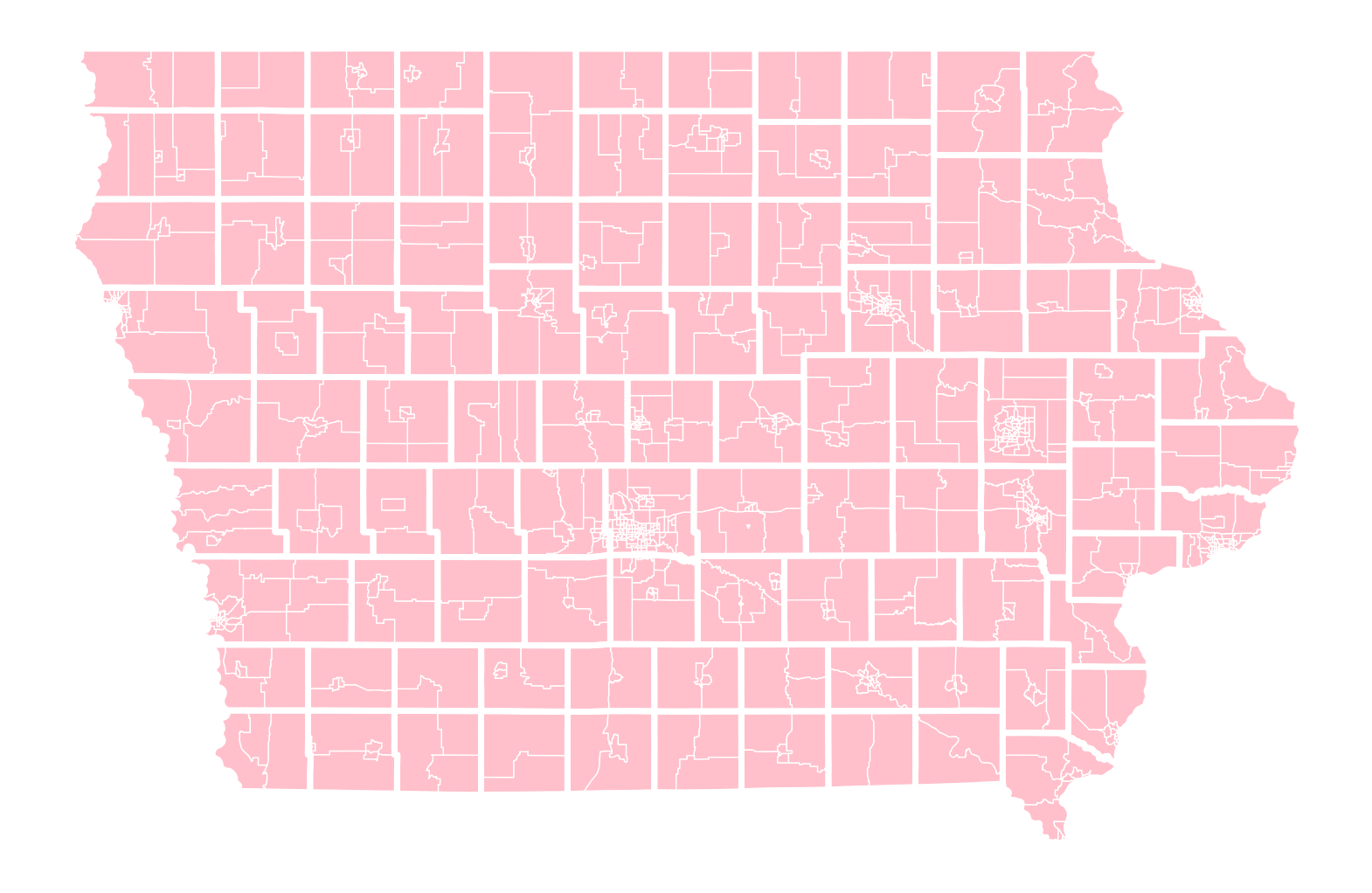}
    \caption{Iowa Counties}
    \label{iowadistricts}
    \end{center}
    \end{wrapfigure}

\subsection{Stochastic Rebalancing of KMeans Generated Districts}

This section describes the process of adjusting the regions to meet population constraints while ensuring that the districts maintained a minimum level of compactness. The algorithm iteratively reassigned counties to achieve a balanced population distribution among districts, subject to compactness constraints.

\subsubsection{KMeans}
The first step in the districting process involved the use of KMeans clustering to generate an initial districting plan based on the population distribution and geographic location of counties. 

Initially, we calculated the centroids of the counties and used as the basis for clustering. The centroids' x and y coordinates were extracted and stored in an array X, and the population values of the counties were stored in the weights array. This setup allowed the clustering algorithm to consider both the geographic location and the population size of the counties.

We then applied the KMeans clustering algorithm to this data, with the number of clusters set to $N = 4$, which corresponded to the desired number of districts. The ${sample\_weight}$ parameter of the KMeans algorithm was used to weight the clustering by the population of each county, ensuring that the resulting districts were balanced in terms of population. After fitting the KMeans model to the data, the labels assigned by the clustering algorithm were added to the GeoDataFrame as the ``district'' column, indicating the district assignment for each county.

This initial districting served as a starting point for further refinement through the subsequent steps of the MCMC simulation, which adjusted the districts to meet additional criteria for compactness and contiguity.

\subsubsection{Adjusting Regions to Meet Population Constraints}

The iteration loop was set up with a maximum of 1000 iterations to adjust the population distribution among districts. This loop was run three times before it found a configuration that met our population and compactness needs. At each step, it chose a plan that improved upon the previous one until it reached one that met the minimum requirements. In each iteration, the total population for each district was calculated using the ``group by'' function on the "district" column and summing the "population" values. Additionally, the minimum Polsby-Popper compactness score across all districts was computed to measure how compact the districts were geometrically. The algorithm then identified districts that were underpopulated (population below the minimum threshold) and overpopulated (population above the maximum threshold). If there were no underpopulated or overpopulated districts, each district was contiguous, and the minimum compactness was above 95\% of the original minimum compactness, the loop broke, indicating that the districts were acceptable.

If there were underpopulated districts, a random underpopulated district was selected. A county was randomly chosen from any overpopulated district and reassigned to the underpopulated district. The new district assignment was then checked for acceptability in terms of contiguity and compactness. If the assignment was acceptable, it was retained; otherwise, it was reverted. If there were no underpopulated districts, the algorithm tried to balance populations within overpopulated districts. A county from an overpopulated district was randomly selected and reassigned to a potential underpopulated district, provided it did not exceed the maximum population for that district. The new assignment was checked for acceptability in terms of contiguity and compactness as well. If the assignment was acceptable, it was retained; otherwise, it was reverted.

In fewer than 3000 iterations a balanced, contiguous, and compact configuration was found, the loop terminated. The final populations of the districts were checked to ensure they were within the specified tolerance of the target population. The resulting districts were then visualized, coloring each district differently for clarity. This approach ensured that the districts were balanced in terms of population and maintained a reasonable level of compactness. The use of random selection and reassignment, combined with contiguity and compactness checks, helped in exploring different configurations efficiently while adhering to the constraints.

\subsection{Monte Carlo Markov Chain Simulation}

As we saw earlier, the initial state of the districts was set using the KMeans rebalanced plan, ensuring that each district adhered to population balance, contiguity, and compactness criteria. At each iteration, the algorithm generated a new districting plan by randomly selecting a county and reassigning it to a different district. The proposed districting plan underwent evaluation, which checked for population balance, contiguity, and compactness. If the proposed plan satisfied all the acceptance criteria, it became the current districting configuration. Otherwise, the plan was rejected, and the previous configuration was retained. The number of edges cut by the new districting plan was then counted and logged.

\subsection{Gaussian Mixture Model Analysis}

To better understand the distribution of cut edges in the generated districting plans, we employed a Gaussian Mixture Model (GMM) analysis. The GMM is a probabilistic model that assumes all data points are generated from a mixture of several Gaussian distributions with unknown parameters.

\begin{definition}[Gaussian Mixtures] 
A \textbf{Gaussian Mixture Model (GMM)} is a weighted sum of $K$ Gaussian component densities, given by the formula:
\begin{equation}
p(x) = \sum_{k=1}^{K} \phi_k \cdot \mathcal{N}(x \mid \mu_k, \sigma_k^2)
\end{equation}
where $\phi_k$ represents the weight of the $k$-th Gaussian component, $\mu_k$ is the mean, and $\sigma_k^2$ is the variance of the $k$-th Gaussian component.
\end{definition} \vspace{10pt}

\noindent By applying the GMM, we assessed the likelihood of the official Iowa plan.

\section{Results}

\subsection{Known Variables}
The following variables are known and fixed for the state of Iowa:

\begin{itemize}
    \item $Population_{USA} = 334,994,511$
    \item $Population_{Iowa} = 3,203,345$
\end{itemize}

We can now calculate the number of seats Iowa gets:

\begin{align}
N &\approx 435 \times \frac{Population_{Iowa}}{Population_{USA}} \\
    &= 435 \times \frac{3,203,345}{334,994,511} \\
    &\approx 4
\end{align} 

\noindent Based on the state population, Iowa gets four seats in the House of Representatives, leading to four districts. For $i \in \{1, \ldots, N\}$ we then have:

\begin{align}
\|\delta_{i}\| &\approx \frac{Population_{Iowa}}{N} \\
                   &= 435 \times \frac{3,203,345}{4} \\
                   &\approx 797,592
\end{align} 

\subsection{Parameters}
The following parameters are adjustable and have been chosen for this study:
\begin{itemize}

\item $\epsilon = 14 \text{mi} \approx 22.556 \text{km}$

\item $\theta = 5\%$

\item $\kappa = 5\%$

\end{itemize}

\subsection{KMeans Clustering for Initial Districts}
\begin{figure}[h]
    \begin{center}
     \captionsetup{justification=centering}
    \includegraphics[width=\linewidth]{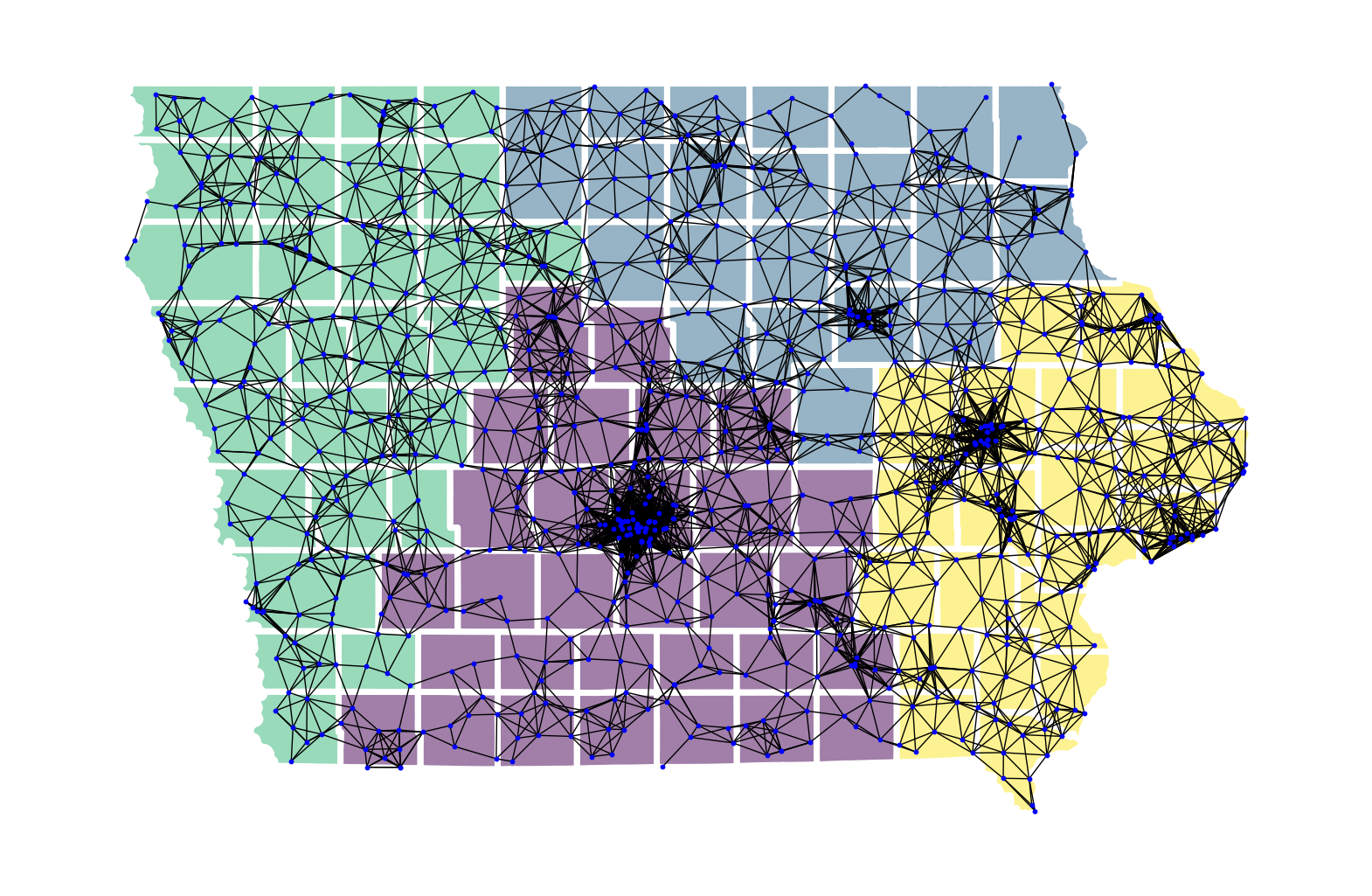}
    \caption{Iowa KMeans Generated Districts with Postal Network. Total Cut edges: 276.}
    \label{iowakmeans}
    \end{center}
\end{figure}

Running KMeans on the county maps from Iowa, with each county represented by its centroid weighted by population size, we obtained the initial districting map shown in Figure \ref{iowakmeans}.

\begin{table}[h]
    \centering
    \begin{tabular}{|c|c|c|c|}
        \hline
        \textbf{District} & \textbf{Population} & \textbf{Population Deviation (\%)} & \textbf{Polsby Popper} \\ \hline
        0 & 1,169,098 & { }46.58 & 0.44 \\ \hline
        1 & 465,563 & -41.63 & 0.42 \\ \hline
        2 & 531,316 & -33.39 & 0.30 \\ \hline
        3 & 1,024,392 & { }28.44 & 0.51 \\ \hline
    \end{tabular}
    \caption{KMeans Districts Info}
    \label{table:iowakmeans}
\end{table}

\noindent The initial KMeans clustering shows significant deviations in population balance, with deviations ranging from -41.63\% to 46.58\%. While the compactness, as indicated by the Polsby-Popper scores, is relatively better, the high population deviation makes this initial plan impractical for fair representation.

\subsection{Rebalancing KMeans Districts}

\begin{figure}[h]
    \begin{center}
     \captionsetup{justification=centering}
    \includegraphics[width=\linewidth]{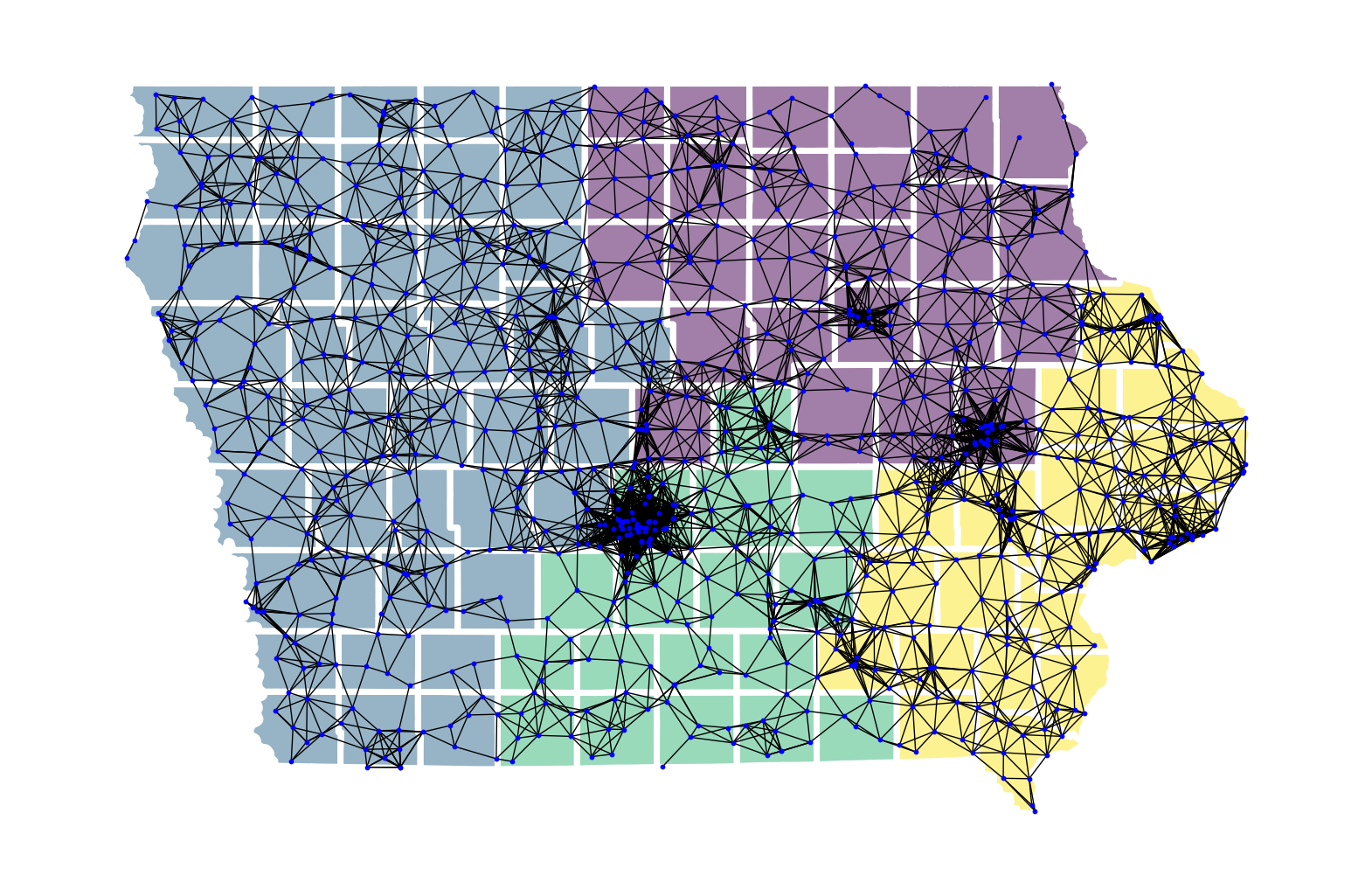}
    \caption{Iowa KMeans Generated Districts Rebalanced with Postal Network. Total Cut edges: 553.}
    \label{iowakmeansrebalance}
    \end{center}
\end{figure}

After rebalancing the KMeans-generated districts to meet population balance, contiguity, and compactness criteria, the resulting districting map is shown in Figure \ref{iowakmeansrebalance}. This rebalanced map significantly reduces cut edges while maintaining better population balance and compactness.

\begin{table}[h]
    \centering
    \begin{tabular}{|c|c|c|c|}
        \hline
        \textbf{District} & \textbf{Population} & \textbf{Population Deviation (\%)} & \textbf{Polsby Popper} \\ \hline
        0 & 822,634 & { }3.14 & 0.46 \\ \hline
        1 & 789,403 & -1.03 & 0.41 \\ \hline
        2 & 791,865 & -0.72 & 0.43 \\ \hline
        3 & 786,467 & -1.39 & 0.39 \\ \hline
    \end{tabular}
    \caption{Districts Info After Rebalancing KMeans }
    \label{table:iowakmeansrebalance}
\end{table}

\noindent The rebalanced KMeans districts show significant improvements in population balance, with deviations ranging from -1.39\% to 3.14\%. The Polsby-Popper scores indicate good compactness across all districts. This rebalanced plan performs better than the initial KMeans plan and is comparable to the official map in terms of compactness, as shown in Table \ref{table:iowakmeansrebalance}.

\subsection{Comparison with Official District Plan}

The official Iowa districting plan, shown in Table \ref{table:district_population}, has excellent population balance, with deviations ranging from -0.0052\% to 0.0066\%. However, its compactness, as measured by the Polsby-Popper score, is relatively lower, ranging from 0.26 to 0.39.

After rebalancing, the resulting districting map has fewer cut edges compared to the official districting map (shown in Figure \ref{iowa_districts_network}), and a minimum Polsby-Popper score among its districts higher than the official map. The only measure where the original map performed better was population balance.

  \begin{figure}[h]
    \begin{center}
     \captionsetup{justification=centering}
    \includegraphics[width=\linewidth]{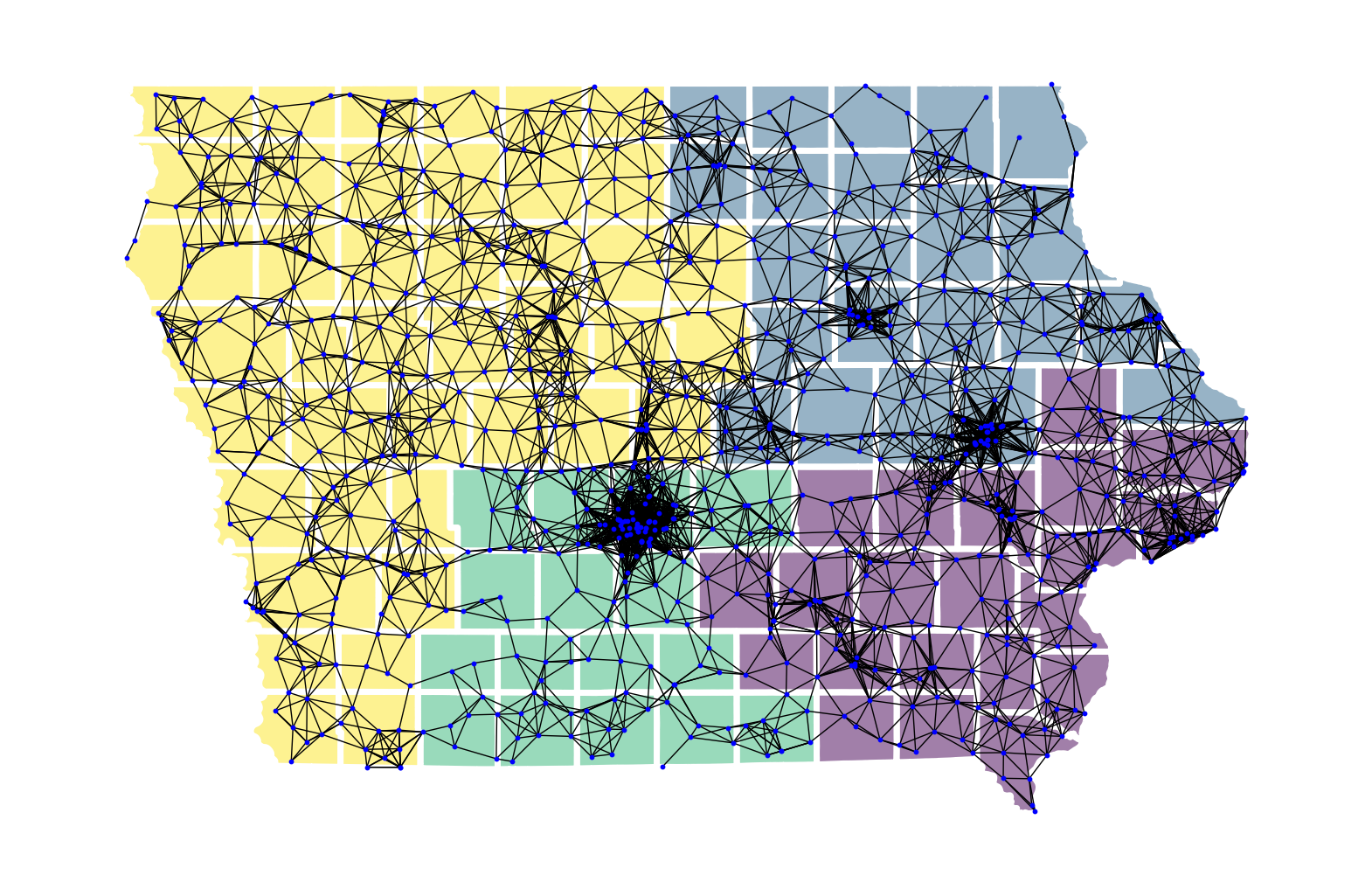}
    \caption{Iowa Minimum Cut Edge plan Observed. Total Cut edges: 315.}
    \end{center}
\end{figure}

\begin{table}[h]
    \centering
    \begin{tabular}{|c|c|c|c|}
        \hline
        \textbf{District} & \textbf{Population} & \textbf{Population Deviation (\%)} & \textbf{Polsby Popper} \\ \hline
        0 & 758,669 & -4.88 & 0.38 \\ \hline
        1 & 821,639 & { }3.02 & 0.40 \\ \hline
        2 & 788,204 & -1.18 & 0.43 \\ \hline
        3 & 821,857 & { }3.04 & 0.37 \\ \hline
    \end{tabular}
    \caption{Districts in Best Configuration Found by MCMC}
    \label{table:district_population_mcmc}
\end{table}

\noindent The best configuration encountered by the MCMC simulation, shown in Table \ref{table:district_population_mcmc}, has population deviations ranging from -4.88\% to 3.04\%. The Polsby-Popper scores indicate good compactness, with scores ranging from 0.37 to 0.43. This configuration also reduces the number of cut edges, enhancing community cohesion.

\subsection{Further Investigation}
\begin{figure}[h]
    \begin{center}
     \captionsetup{justification=centering}
    \includegraphics[width=\linewidth]{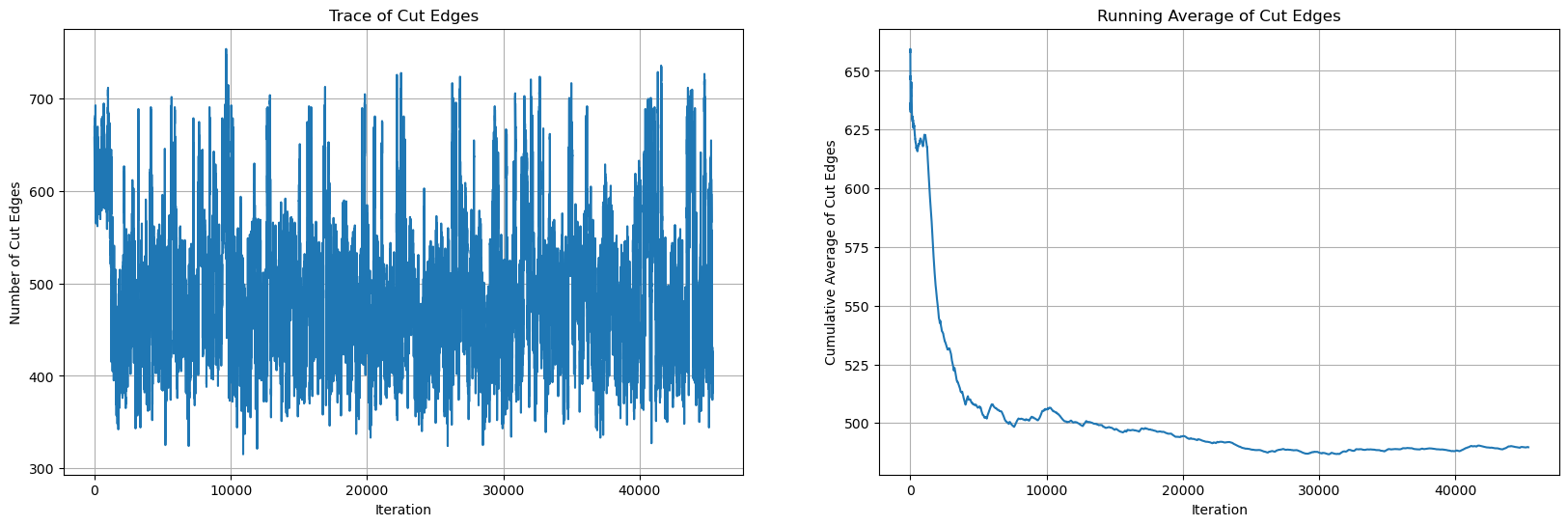}
    \caption{MCMC Trace and Running Average}
    \label{fig:TraceRunningAvg}
    \end{center}
\end{figure}

\begin{figure}[h]
    \begin{center}
     \captionsetup{justification=centering}
    \includegraphics[width=\linewidth]{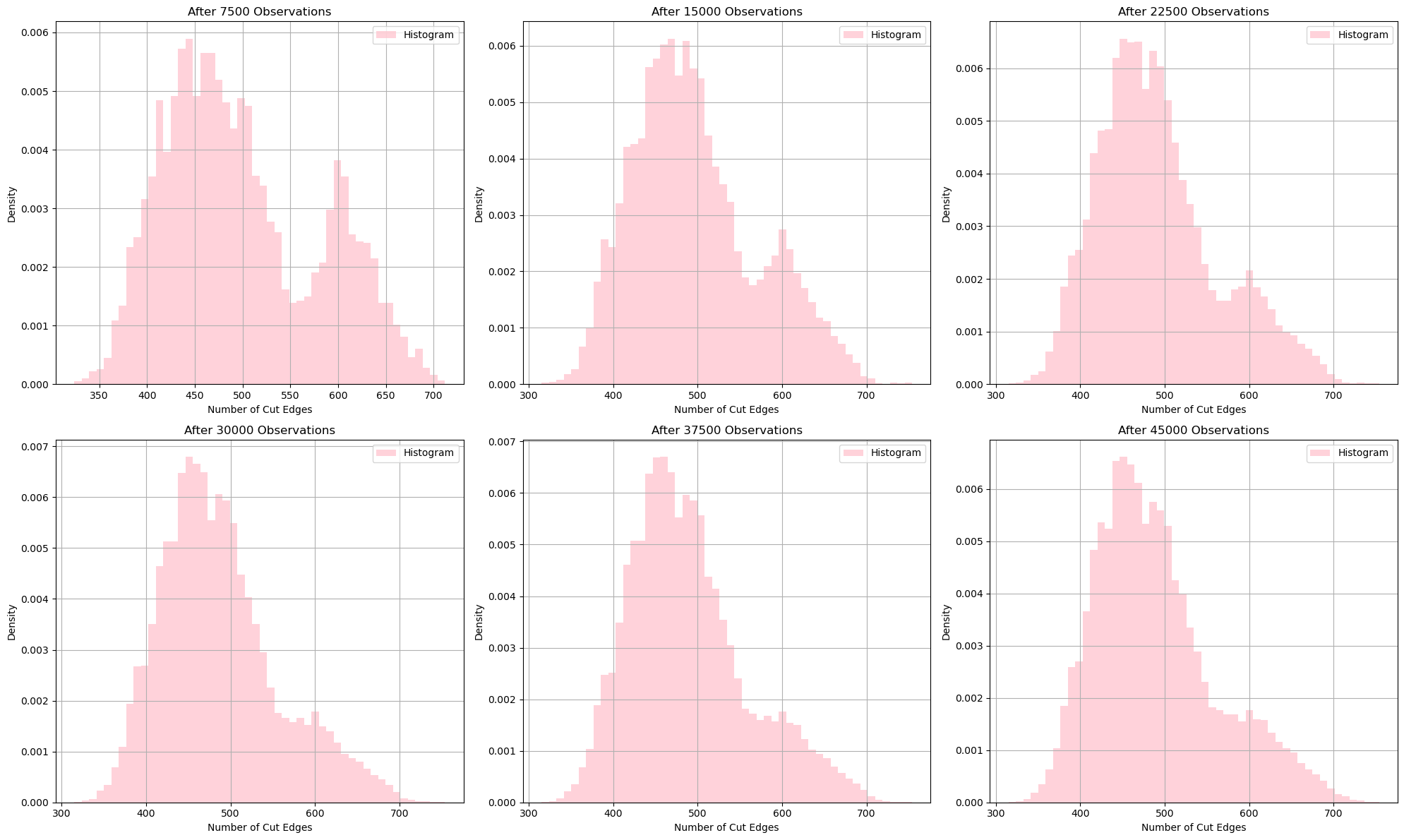}
    \caption{MCMC Observed Plans Cut Edges Distribution}
     \label{fig:MCMCRunsNoNormal}
    \end{center}
\end{figure}

We observed that, although the running average had stabilized rather quickly, the trace plots exhibited periodic behavior (Figure \ref{fig:TraceRunningAvg}) and persistent bumps in the histogram that didn't quite dissappear over time (Figure \ref{fig:MCMCRunsNoNormal}). This led us to investigate the possibility that the observed data might originate from multiple distributions.

\noindent We performed BIC analysis to find the optimal number of Gaussian Distributions needed for a Gaussian Mixture to effectively model the data collected and obtained $n = 3$. See Figure \ref{fig:BICGaussians} below.

 \begin{figure}[h]
    \begin{center}
     \captionsetup{justification=centering}
    \includegraphics[width=.6\linewidth]{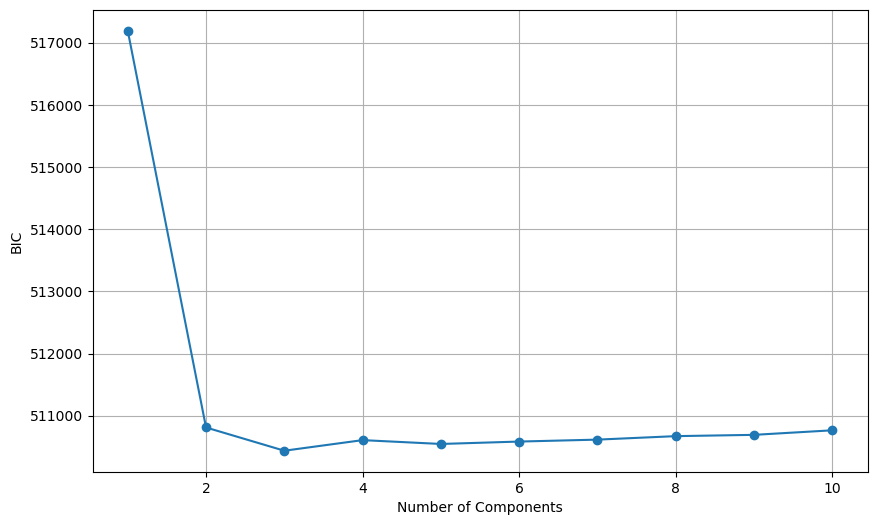}
    \caption{BIC for Different Numbers of Gaussian Components}
    \label{fig:BICGaussians}
    \end{center}
\end{figure}

\begin{table}[h]
\centering
\begin{tabular}{|c|c|c|c|}
\hline
\textbf{Component} & \textbf{Mean} & \textbf{Variance} & \textbf{Weight} \\ \hline
1 & 498.37 & 1013.28 & 0.38466576 \\ \hline
2 & 605.86 & 2047.77 & 0.18806542 \\ \hline
3 & 430.81 & 1018.74 & 0.42726882 \\ \hline
\end{tabular}
\caption{Gaussian Mixture Model Parameters}
\label{table:gmm_parameters}
\end{table}



\begin{figure}[h]
    \begin{center}
     \captionsetup{justification=centering}
    \includegraphics[width=.8\linewidth]{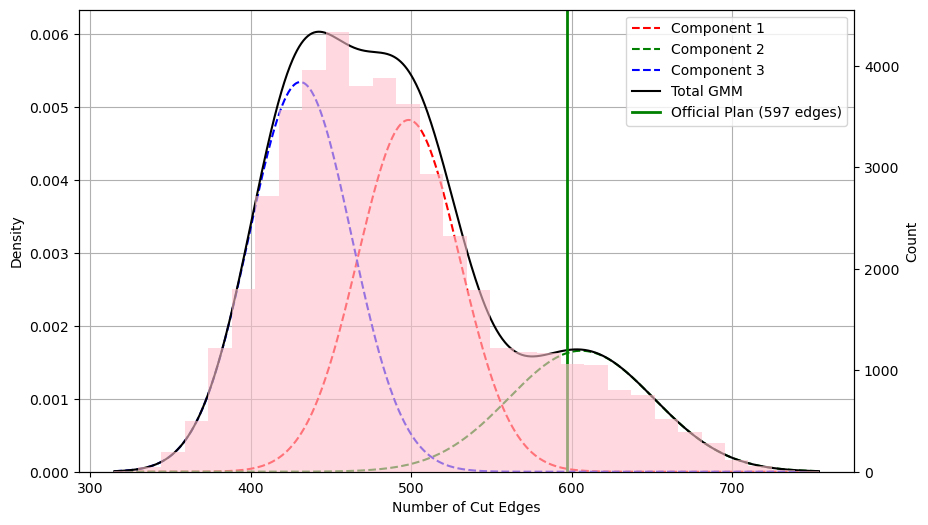}
    \caption{Histogram with Gaussian Mixture Model}
     \label{fig:MCMCTotalwOG}
    \end{center}
\end{figure}

Based on the Gaussian Mixture Model (GMM) analysis, the cumulative probability \( P(Z < 597) \) is 0.8910, meaning that 89.10\% of the generated districting plans have fewer cut edges than the official Iowa districting plan. This result is derived using the formula:

\[
P(Z < 597) = \sum_{k=1}^{3} \phi_k \cdot \left( \int_{-\infty}^{597} \frac{1}{\sqrt{2\pi \sigma_k^2}} \exp\left( -\frac{(x - \mu_k)^2}{2\sigma_k^2} \right) dx \right)
\]

\noindent where $\phi_k$, $\mu_k$, and $\sigma_k^2$ are the weight, mean, and variance of the $k$-th Gaussian component found in Table \ref{table:gmm_parameters}, respectively. This high cumulative probability suggests that the official districting plan is an outlier, as the majority of the generated plans exhibit fewer disruptions to community cohesion. Consequently, if a districting plan were chosen at random from our generated distribution, there is an 89\% chance that it would have fewer cut edges than the current official plan. This indicates potential inefficiencies or irregularities in how the official plan was drawn, warranting further investigation for fairness and compliance with redistricting principles. See Figure \ref{fig:MCMCTotalwOG} for comparison.

  \begin{figure}[h]
    \begin{center}
     \captionsetup{justification=centering}
    \includegraphics[width=\linewidth]{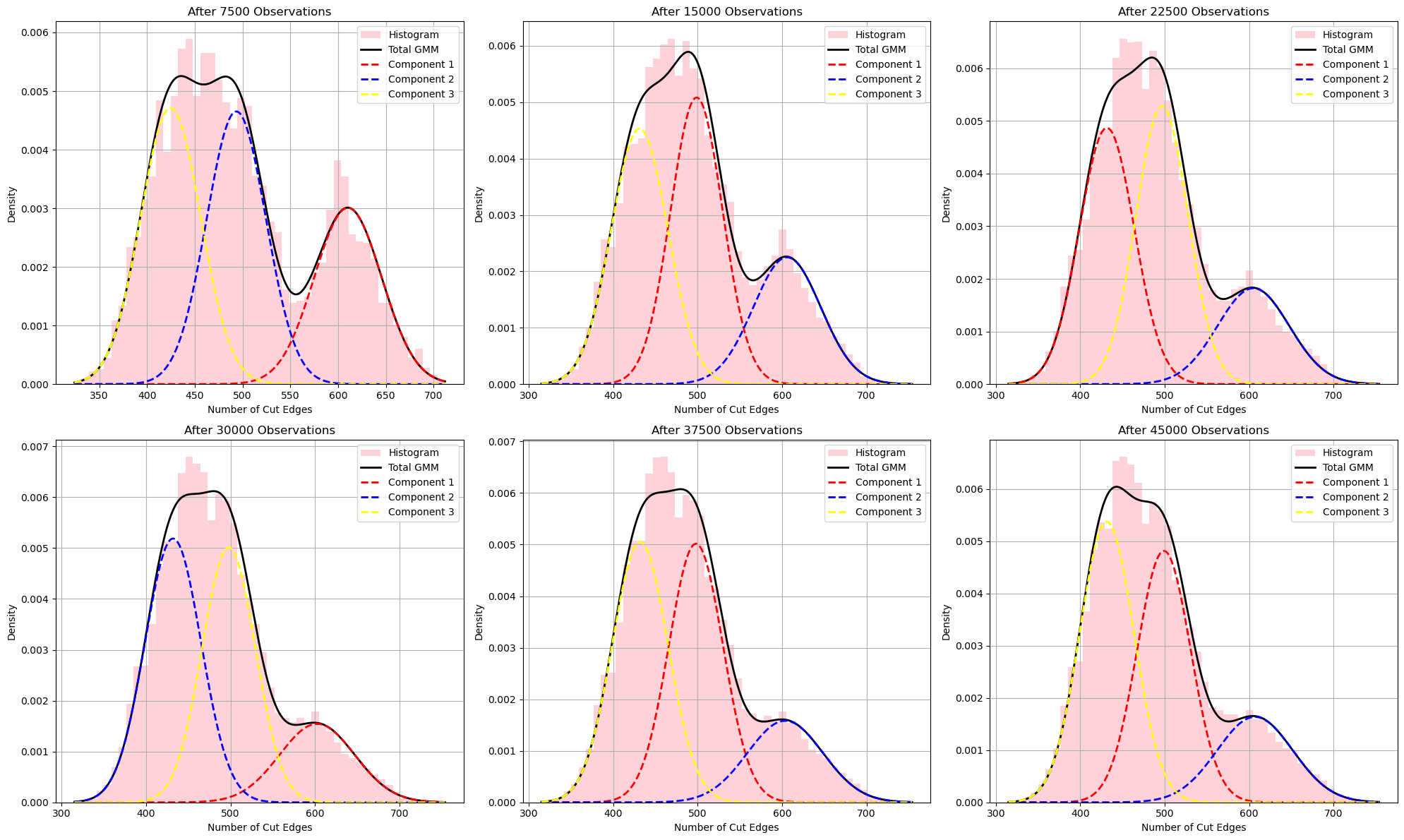}
    \caption{MCMC Observed Plans Histogram Fitted}
     \label{fig:MCMCRunsGaussian3Fitted}
    \end{center}
\end{figure}

\section{Discussion}

\subsection{Key Findings}
\begin{enumerate}
    \item \textbf{Effectiveness of TDA and Postal Network:} \\ 
    Our method enhances traditional redistricting practices by using post offices as population hubs, providing a more accurate representation of population distribution and community cohesion. By leveraging persistent homology within TDA, we gain detailed insights into spatial distributions and connections, allowing for a sophisticated evaluation of community integrity that surpasses traditional centroid-based measures.

    \item \textbf{Performance of Rebalanced Districts:} \\ 
    The rebalanced districting plans generated using our methodology produced fewer cut edges and better compactness scores compared to the official districting plan for Iowa. However, the official plan performed better in terms of population balance. This highlights the trade-offs between different redistricting criteria.

    \item \textbf{Probabilistic Evaluation:} \\ 
    Our MCMC simulation revealed that the likelihood of encountering a districting plan similar to the official Iowa plan within our generated distribution is exceedingly low. This suggests potential inefficiencies or irregularities in the official plan, warranting further investigation for fairness and compliance with redistricting principles.
\end{enumerate}

\subsection{Limitations}
\begin{enumerate}
   \item \textbf{Data and Computational Constraints:} \\
   The study relies on data from the U.S. Census Bureau and the USPS network. Any inaccuracies or outdated information in these datasets could impact the results. Additionally, the computational demands of the MCMC simulations and TDA are substantial. For this study, out of 990,000 iterations (after a burn-in period of 10,000), only 45,344 valid plans were encountered. This process took approximately an hour per 10,000 iterations on average. Although the computations were feasible for Iowa, scaling this approach to larger states or using finer granularity data, such as census tracts, would require significant computational resources. The high computational cost limits the practicality of the approach for real-time or large-scale applications without access to substantial computing power. Future work could explore integration with libraries such as ReCom and GerryChain created by MGGG (Metric Geometry and Gerrymandering Group) and discussed by DeFord et al. \cite{deford2019mcmc} to make the process to make our sampling more efficient while also contributing TDA cut edges to their package.

  \item \textbf{Parameter Sensitivity:} \\
  The results are sensitive to the choice of parameters, such as the percentile for TDA ($p$), the population deviation threshold ($\theta$), and the compactness relaxation factor ($\kappa$). Different choices for these parameters can lead to different districting plans, and there is no universally accepted method for selecting optimal values. This sensitivity introduces an element of subjectivity into the process. Investigating the optimal $p$ percentile for TDA and understanding the implications of different filtrations can significantly enhance the robustness of the analysis.

    \item \textbf{Generalizability:} \\
    While Iowa was used as a case study due to its requirement to make districts respect county lines, simplifying computations, this approach could be extended to the Census tract level with more computational power. Exploring the use of other services as community proxies could also provide additional insights into community cohesion.
\end{enumerate}

\subsection{Future Work}
Future work could explore the application of this methodology to other states and incorporate additional community proxies beyond the postal network. Further optimizing the computational efficiency of the simulations is also crucial. Investigating the impact of different parameter choices and filtrations in TDA will provide a deeper understanding of the underlying community structures and improve the robustness of the districting plans.

\subsection{Conclusions}

This study presents a new approach to evaluating electoral districting plans by integrating Topological Data Analysis (TDA) with Markov Chain Monte Carlo (MCMC) simulations. Using Iowa as a case study, we demonstrate how the postal network, a proxy for community cohesion, can be utilized to assess the integrity of districting plans. Our methodology leverages KMeans clustering to generate initial districts, followed by a stochastic rebalancing process to ensure population balance, contiguity, and compactness.

The results show that the rebalanced districting plans produced fewer cut edges and achieved better compactness scores compared to the official districting plan, although the official plan performed better in terms of population balance. The MCMC simulation further reveals that the likelihood of encountering a plan similar to the official Iowa districting plan within our generated distribution is exceedingly low, suggesting potential inefficiencies or irregularities in the official plan.

Despite the challenges of data accuracy, computational constraints, and scalability, our approach provides a robust framework for generating and evaluating districting plans. Future work will focus on extending this methodology to other states, incorporating additional community proxies, and optimizing computational efficiency. By refining these methods, we aim to contribute to more equitable and transparent redistricting processes that better reflect community structures and foster fair representation.

\clearpage

\appendix

\section*{Appendix}

\subsection*{Stochastic Rebalancing of KMeans Generated Districts}



\begin{algorithm}[H]
\caption{Accept Interim Proposal}
\begin{algorithmic}[1]
\State \textbf{Input:} Proposed districts, minimum acceptable compactness score
\State \textbf{Output:} Boolean value indicating whether the proposal is accepted
\State Calculate the compactness score for each proposed district
\State Determine the minimum compactness score among the proposed districts
\If{the minimum compactness score is less than the minimum acceptable compactness score}
    \State \textbf{Return} False
\EndIf
\For{each geometry in the proposed districts}
    \If{the geometry is not a simple, valid polygon}
        \State \textbf{Return} False
    \EndIf
\EndFor
\State \textbf{Return} True
\end{algorithmic}
\end{algorithm}

\begin{algorithm}[H]
\caption{Stochastic Rebalancing of KMeans Clusters (SRKMeans)}
\begin{algorithmic}[1]
\State \textbf{Input:} Initial districts, population thresholds (minimum and maximum), original minimum compactness score
\State \textbf{Output:} Balanced districts with populations within tolerance
\State Initialize iteration counter to 0 and set the maximum number of iterations to 1000
\While{iteration counter is less than the maximum number of iterations}
    \State Calculate the population of each district
    \State Calculate the minimum Polsby-Popper compactness score for the current districts
    \State Identify districts that are underpopulated and overpopulated
    \If{there are no underpopulated or overpopulated districts and the minimum compactness score is acceptable}
        \State \textbf{break}
    \EndIf
    \If{there are underpopulated districts}
        \State Select a random underpopulated district
        \State Sample a county from any overpopulated district
        \State Move the sampled county to the selected underpopulated district
        \State Calculate the updated district geometries
        \If{the new district configuration is acceptable}
            \State \textbf{continue}
        \Else
            \State Revert the county move
        \EndIf
    \Else
        \For{each overpopulated district}
            \State Sample a county from the overpopulated district
            \For{each district below the target population}
                \If{moving the sampled county does not exceed the maximum population threshold}
                    \State Move the county to the selected district
                    \State Calculate the updated district geometries
                    \If{the new district configuration is acceptable}
                        \State \textbf{break}
                    \Else
                        \State Revert the county move
                    \EndIf
                \EndIf
            \EndFor
        \EndFor
    \EndIf
    \State Increment the iteration counter
\EndWhile
\State \textbf{Output:} Final district populations and compactness within tolerance
\end{algorithmic}
\end{algorithm}

\subsection*{Proposal Generation}

\begin{algorithm}[H]
\caption{Propose New Districts}
\begin{algorithmic}[1]
\State \textbf{Input:} Geospatial dataframe of current counties
\State \textbf{Output:} Proposed districts
\State Randomly select a county and reassign it to a different district
\State Create new proposed districts based on the modified assignments
\State \textbf{Return} the proposed districts
\end{algorithmic}
\end{algorithm}

\subsection*{Acceptance Criteria}

The proposed districting plan is evaluated using the function \texttt{accept\_proposal}, which checks for the following:

\begin{itemize}
    \item \textbf{Population Balance:} Each district's population must be within a specified deviation threshold from the average district population.
    \item \textbf{Contiguity:} Each district must form a single contiguous region.
    \item \textbf{Compactness:} Each district must meet a minimum compactness criterion defined by the Polsby-Popper measure.
\end{itemize}

A plan is accepted if it satisfies these criteria. The detailed function for checking these conditions is as follows:

\begin{algorithm}[H]
\caption{Accept Proposal}
\begin{algorithmic}[1]
\State \textbf{Input:} Proposed districts, minimum compactness score, compactness threshold
\If{the population of each district is within acceptable limits}
    \If{the compactness score of each district is above the threshold}
        \For{each district geometry}
            \If{the geometry is contiguous and valid}
                \State \textbf{Accept} the proposal
            \Else
                \State \textbf{Reject} the proposal
            \EndIf
        \EndFor
    \Else
        \State \textbf{Reject} the proposal
    \EndIf
\Else
    \State \textbf{Reject} the proposal
\EndIf
\end{algorithmic}
\end{algorithm}

\subsection*{Recording Results}

\begin{algorithm}[H]
\caption{Calculate Cut Edges}
\begin{algorithmic}[1]
\State \textbf{Input:} Districts, geospatial edges
\State \textbf{Output:} Number of cut edges
\State Initialize the cut edges counter to 0
\For{each edge in the geospatial edges}
    \If{the edge is completely within a district}
        \State Skip to the next edge
    \Else
        \State Check intersection with district boundaries
        \If{the edge intersects boundaries and is not identical to the boundary}
            \State Increment the cut edges counter
        \EndIf
    \EndIf
\EndFor
\State \textbf{Return} the number of cut edges
\end{algorithmic}
\end{algorithm}


\begin{algorithm}[H]
\caption{MCMC Simulation}
\begin{algorithmic}[1]
\State \textbf{Input:} Geospatial dataframe of counties, geospatial edges dataframe, burn-in rate, number of iterations
\State Initialize the burn-in period as the product of the number of iterations and the burn-in rate
\State Set the initial state as the result of generating initial districts from the geospatial dataframe of counties
\State Initialize an empty history array
\State Set the minimum number of cut edges to infinity
\State Initialize the best configuration as None
\For{each iteration from 1 to the number of iterations}
    \State Propose new districts from the current districts
    \If{the proposed districts are accepted}
        \State Update the current districts to the proposed districts
        \If{the current iteration is greater than or equal to the burn-in period}
            \State Calculate the number of cut edges in the current districts
            \State Append the number of cut edges to the history array
            \If{the number of cut edges is less than the minimum number of cut edges}
                \State Update the minimum number of cut edges and the best configuration
            \EndIf
        \EndIf
    \Else
        \If{the current iteration is greater than or equal to the burn-in period}
            \State Append None to the history array
        \EndIf
    \EndIf
\EndFor
\State \textbf{Return} the history array, the minimum number of cut edges, and the best configuration
\end{algorithmic}
\end{algorithm}

\subsection*{Analyzing Results}

\begin{algorithm}[H]
\caption{Analyze Results}
\begin{algorithmic}[1]
\State \textbf{Input:} History of cut edges, cut edges count of the original plan
\State Fit a normal or skewed normal distribution to the history of cut edges
\State Calculate the p-value for choosing a plan worse than the original plan
\State \textbf{Return} the mean, variance, and p-value
\end{algorithmic}
\end{algorithm}

\end{document}